\newif\ifsingle
\singlefalse 

\ifsingle
\documentclass[11pt,draftclsnofoot, onecolumn]{IEEEtran}		
\else		
\documentclass[10pt,final, twocolumn]{IEEEtran}
\fi

\usepackage{times}
\usepackage{amsmath,dsfont}
\usepackage{amssymb,amsthm}
\usepackage{epsfig,verbatim}
\usepackage{subfigure}
\usepackage{setspace}
\usepackage{color}
\usepackage{cite}
\usepackage{epstopdf}
\usepackage{graphicx}

\usepackage{accents}
\usepackage{acronym}
\usepackage[bookmarks,colorlinks]{hyperref}
\usepackage{booktabs}
\usepackage{mathtools}
\usepackage{enumitem}
\usepackage{framed}

\usepackage{float}

\usepackage{algpseudocode} 

\usepackage[ruled,linesnumbered]{algorithm2e}  

\newcommand{\myVec}[1]{{\mathbf{#1}}}
\newcommand{\myMat}[1]{{\mathbf{#1}}}
\newcommand{\mySet}[1]{\mathcal{#1}}
\newcommand{\E}{\mathds{E}}		 			
\newcommand{\myY}{\myVec{y}} 
\newcommand{\myParams}{\myVec{\theta}}

\newcommand{\Nusers}{K}
\newcommand{\Nantennas}{N}
 
\newcommand{\Ncoeffs}{P}
\newcommand{\Ntraining}{T}

\newcommand{\Objective}{\mathcal{F}}
\newcommand{\EmpObjective}{F}

\definecolor{NewColor}{rgb}{0,0,0} 

\ifsingle

\newcommand{\includefig}[1]{\includegraphics[width = 0.75\columnwidth]{#1}} 
\else

\newcommand{\includefig}[1]{\includegraphics[width = 0.95\columnwidth]{#1}} 	

\fi 

\acrodef{adc}[ADC]{analog-to-digital convertor}
\acrodef{cs}[CS]{compressed sensing}
\acrodef{dtft}[DTFT]{discrete-time Fourier transform}
\acrodef{dnn}[DNN]{Deep Neural Network} 
\acrodef{csi}[CSI]{Channel State Information}
\acrodef{map}[MAP]{Maximum A-posteriori Probability}
\acrodef{snr}[SNR]{Signal-to-Noise Ratio}
\acrodef{bs}[BS]{Base Station} 
\acrodef{em}[EM]{electromagnetic} 
\acrodef{iot}[IOT]{Interent of Things}
\acrodef{mimo}[MIMO]{Multiple-Input Multiple-Output}
\acrodef{mse}[MSE]{mean-squared error}
\acrodef{pdf}[PDF]{probability density function}
\acrodef{rv}[RV]{random variable}
\acrodef{ml}[ML]{Machine Learning}
\acrodef{fec}[FEC]{forward error correction}
\acrodef{rs}[RS]{Reed-Solomon}
\acrodef{lti}[LTI]{linear time-invariant}
\acrodef{wss}[WSS]{wide-sense stationary}
\acrodef{psd}[PSD]{power spectral density}
\acrodef{ser}[SER]{symbol error rate} 
\acrodef{ber}[BER]{Bit Error Rate} 
\acrodef{sgd}[SGD]{stochastic gradient descent} 
\acrodef{isi}[ISI]{intersymbol interference}  
\acrodef{awgn}[AWGN]{additive white Gaussian noise} 
\acrodef{ut}[UT]{User Terminal} 
\acrodef{mmw}[mmWave]{millimeter wave}
\acrodef{noma}[NOMA]{non-orthognal multiple access}
\acrodef{mac}[MAC]{mulitple access channel}
\acrodef{fl}[FL]{Federated learning}
\acrodef{ris}[RIS]{Reconfigurable Intelligent Surface} 
\acrodef{deepsic}[DeepSIC]{deep soft inteference cancellation}
\acrodef{bo}[BO]{Bayesian Optimization}
\acrodef{gp}[GP]{Gaussian Process}
\acrodef{ei}[EI]{Expected Improvement}
\acrodef{snr}[SNR]{signal-to-noise ratio}
\IEEEoverridecommandlockouts

\title{Jointly Learned Symbol Detection and Signal Reflection in RIS-Aided Multi-user MIMO Systems
}

\author{
	\IEEEauthorblockN{Liuhang Wang, Nir Shlezinger, George C. Alexandropoulos, \\Haiyang Zhang, Baoyun Wang, and 
	Yonina C. Eldar
	} 
	\thanks{
	L. Wang and B. Wang are with the School of Communication and Information Engineering, Nanjing University of Posts and Telecommunications, Nanjing, China (e-mail:2331829836@qq.com; bywang@njupt.edu.cn).
		N. Shlezinger is with the School of Electrical and Computer Engineering, Ben-Gurion University of the Negev, Beer-Sheva, Israel (e-mail: nirshl@bgu.ac.il).
		G. C. Alexandropoulos is with the Department of Informatics and Telecommunications, National and Kapodistrian University of Athens, Greece (e-mail: alexandg@di.uoa.gr).
		H. Zhang and Y. C. Eldar are with the Faculty of Math and Computer Science, Weizmann Institute of Science, Rehovot, Israel (e-mail: \{haiyang.zhang; yonina\}@weizmann.ac.il). This work has been partially supported by the EU H2020 RISE-6G project under grant number 101017011.}
}

\vspace{-0.75cm}

\begin{document}

	\maketitle
	\pagestyle{plain}
	\thispagestyle{plain}

	\begin{abstract}
 Reconfigurable Intelligent Surfaces (RISs) are regarded as a key technology for future wireless communications, enabling programmable radio propagation environments. However, the passive   reflecting feature of RISs induces notable challenges on channel estimation, making coherent symbol detection a challenging task. In this paper, we consider the uplink of RIS-aided multi-user Multiple-Input Multiple-Output (MIMO) systems and propose a Machine Learning (ML) approach to jointly design the multi-antenna receiver and configure the RIS reflection coefficients, which does not require explicit full knowledge of the channel input-output relationship. Our approach devises a ML-based receiver, while the configurations of the RIS reflection patterns affecting the underlying propagation channel are treated as hyperparameters. Based on this system design formulation, we propose a Bayesian ML framework for optimizing the RIS hyperparameters, according to which the transmitted pilots are directly used to jointly tune the RIS and the multi-antenna receiver. Our simulation results demonstrate the capability of the proposed approach to provide reliable communications in non-linear channel conditions corrupted by Gaussian noise. 

\end{abstract}

	\vspace{-0.4cm}
	\section{Introduction}
	\vspace{-0.1cm} 
	Modern communications systems are subject to constantly growing throughput requirements. In order to meet these demands, \acp{bs} are commonly equipped with multiple antennas, and communicate with several users simultaneously to increase the spectral efficiency \cite{marzetta2010noncooperative}. One of the main challenges in such multi-user \ac{mimo} systems is symbol detection, namely, the recovery of the multiple symbols transmitted over the uplink channel at the \ac{bs}.  Conventional detection algorithms, such as those based on the \ac{map} rule, which jointly recovers all the symbols simultaneously, become infeasible as the number of symbols grows. Alternative low complexity symbol detection are usually based on separate detection or iterative interference cancellation methods \cite{andrews2005interference}, which allow to achieve \ac{map}-approaching performance at complexity which only grows linearly with the number of users. In addition, even when the channel model is linear and known, inaccurate knowledge of the parameters of the channel, namely,  \ac{csi} uncertainty, can significantly degrade the performance. 
	
	An alternative data-driven approach to model-based algorithms uses \ac{ml}. \acp{dnn}, which constitute a popular ML approach, have demonstrated an unprecedented empirical success in various applications, including image and speech processing \cite{lecun2015deep}. 
	Recent years have witnessed growing interest in the application of \acp{dnn} for Receiver (RX) design;  see detailed surveys in \cite{oshea2017introduction,  mao2018deep, gunduz2019machine, balatsoukas2019deep}. 
	Unlike model-based reception, which implements a specified detection rule, \ac{ml}-based RXs learn how to map the channel outputs into the transmitted symbols from training, namely, they operate in a data-driven manner. Multiple \ac{ml}-aided \ac{mimo} reception architectures have been proposed in the literature, including the application of conventional black-box architectures \cite{liao2019deep}, deep unfolded optimization algorithms such as projected gradient descent \cite{samuel2019learning,khobahi2021lord}, and approximate message passing \cite{he2018model}. While the aforementioned RXs involve highly parameterized \acp{dnn}, which require massive volumes of data for training, the more recent work \cite{shlezinger2019deepSIC} designed a data-driven RX which learns to implement the soft iterative interference cancellation symbol detection algorithm of \cite{choi2000iterative} from relatively small labeled data sets. 

	An additional emerging technology for multi-user wireless communication systems is the consideration of \acp{ris} as enablers for controllable signal propagation conditions \cite{huang2019reconfigurable,di2019smart}. This application builds upon the capability of \acp{ris} to generate reconfigurable reflection patterns. An \ac{ris} deployed in urban settings can facilitate and improve communication between the \ac{bs} and multiple users by effectively modifying the propagation of information-bearing signals \cite{WavePropTCCN}. The \acp{ris} enable the communications system as a whole to overcome harsh non Line-Of-Sight (LOS) conditions and improve coverage, when the surface close to the BS or the users, without increasing transmission power. Nonetheless, the fact that \acp{ris} are passive devices, which only reflect their impinging signals in a configurable manner, gives rise to a multitude of signal processing challenges, including complex and costly channel estimation \cite{alexandropoulos2021hybrid}. Furthermore, identifying the proper configuration of the \ac{ris} reflection patterns is a difficult task and requires accurate knowledge of the underlying channel \cite{alexandropoulos2021reconfigurable}, which in turn is quite challenging to acquire. This motivates the application of model-agnostic data-driven \ac{ml} for tunning \acp{ris}, which is the focus here.  
	
	In this paper, considering the uplink of an RIS-empowered multi-user MIMO communication system, we present an ML-based approach to jointly tackle the problem of symbol detection at the BS's multi-antenna RX, along with the configuration of the \ac{ris} reflection coefficients. We adopt the recent DeepSIC reception of \cite{shlezinger2019deepSIC} at the \ac{bs} and treat the \ac{ris} phase configuration as hyperparameters of the learning procedure. We devise an algorithm based on \ac{bo} for the \ac{ris} configuration, which alternatively combined with the \ac{dnn}-aided RX to enable joint optimization of the \ac{ris} and the RX with small amounts for pilot signals. 
	Our numerical results demonstrate the efficacy of the proposed model-agnostic \ac{dnn}-based learning approach for providing reliable RIS-empowered multi-user MIMO communications, without relying on channel modeling and \ac{csi} estimation. 
	
	The rest of this paper is organized as follows. Section \ref{sec:Model} includes the system model and problem formulation, while Section \ref{sec:jointly-learned ris} details the proposed learning approach for jointly learning the multi-antenna RX and the \ac{ris} phase configuration. Section~\ref{sec:simulation study} presents the simulation results, and the paper's conclusions are drawn in Section \ref{sec:conclusions}. Throughout the paper, we use low-case letters for scalars (e.g. $x$), lower case bold-faced letters for vectors (e.g., $\myVec{x}$), upper case bold-faced letters for matrices (e.g., $\myMat{X}$), and calligraphic letters for sets (e.g., $\mathcal{X}$). The $i$-th element of $\myVec{x}$ is denoted by $[\myVec{x}]_i$, $\mathbb{E}[\cdot] $ is the expectation operator, and $ (\cdot)^{\mathrm{T}}$ returns the transpose.
	 
	\vspace{-0.2cm}
	\section{System Model and Design Objective}
	\label{sec:Model}
	\vspace{-0.1cm} 
\subsection{System Model} \label{subsec:system model}
\begin{figure}
	 	\centering
	 	{\includefig{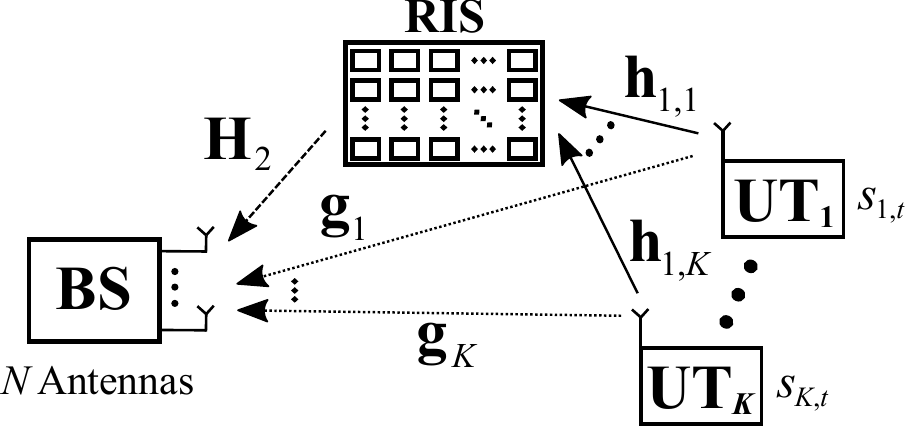}} 
	 	\caption{The considered \ac{ris}-empowered multi-user \ac{mimo} communication system operating in the uplink direction.}
	 	\label{fig:Setup}	 
	    \end{figure}

We focus on the uplink of cellular networks and consider a single-cell including a \ac{bs} with $N$ antenna elements that serves $\Nusers$ User Terminals (UTs), as illustrated in Fig.~\ref{fig:Setup}. This uplink communication is assumed to be assisted by an \ac{ris} with $\Ncoeffs$ unit elements \cite{alexandropoulos2021reconfigurable}. An RIS controller, which is accessible by the BS, handles the metasurface's reflection configuration \cite{RISE6G_COMMAG}. We assume that the \ac{bs} makes use of the \ac{dnn}-based RX in \cite{shlezinger2019deepSIC}, which particularly implements a data-driven detector. In order to train the \ac{bs}, in each time instance $t$, the UTs transmit known pilot symbols, denoted $\myVec{s}_t\triangleq[s_{1,t}\,s_{2,t}\,\cdots\,s_{K,t}] \in \mySet{S}^{\Nusers\times1}$, where $\mySet{S}$ is a discrete constellation set of size $M$.

We assume that the input-output relationship of the wireless channel is given by some stochastic transformation parameterized by the \ac{ris} phase configuration vector $\boldsymbol{\phi}\in\mySet{C}^\Ncoeffs$. Considering finite resolution phase shifting values for the RIS unit elements, each $n$-th element (with $n=1,2,\ldots,\Ncoeffs$) of $\boldsymbol{\phi}$ can be modeled as follows \cite{Huang_GLOBECOM_2019}:
\begin{equation}\label{phase shifting_model}
[\boldsymbol{\phi}]_n \in \left\{e^{j2^{1-b}\pi m}\right\}_{m=0}^{2^b-1},
\end{equation}
where $b$ is the phase resolution in number of bits; clearly, the different number of phase shifting values per RIS unit element is $2^b$. Based on the latter expression, we represent the feasible set of \ac{ris} phase configuration vectors as $\Theta$, i.e., $\boldsymbol{\phi}\in \Theta \subset \mySet{C}^{\Ncoeffs\times1}$. The channel output, i.e., the baseband received at the $\Nantennas$ \ac{bs} antenna elements, at the time instance $t$ is modeled as 	 
\begin{equation}\label{eq:channel_output}
	 \myY_t = f_{\boldsymbol{\phi}}\left( \myVec{s}_t\right) \in \mySet{C}^{\Nantennas},
\end{equation}   
where $f_{\boldsymbol{\phi}}\left(\cdot\right)$ represents an unknown generic function that depends on $\boldsymbol{\phi}$. For the special case of the conventional linear Gaussian channels, this function takes the following form \cite{huang2019reconfigurable}: 
\begin{equation}\label{eq:channel model}
	 f_{\boldsymbol{\phi}}^{\rm (G)}\left( \myVec{s}_t\right) \triangleq \left( \mathbf{H}_2 \boldsymbol{\Phi} \mathbf{H}_1 + \mathbf{G}\right) \myVec{s}_t + \mathbf{n}_t,
\end{equation}   
where $\boldsymbol{\Phi}\triangleq{\rm diag}(\boldsymbol{\phi})$, $\mathbf{H}_1\triangleq[\mathbf{h}_{1,1}\,\mathbf{h}_{1,2}\,\cdots\,\mathbf{h}_{1,K}]\in\mySet{C}^{\Ncoeffs\times\Nusers}$ denotes the wireless channel gain matrix between the RIS unit elements and the UTs ($\mathbf{h}_{1,k}\in\mySet{C}^{\Ncoeffs\times1}$ with $k=1,2,\ldots,K$ represents the channel for the $k$-th UT), $\mathbf{H}_2\in\mySet{C}^{\Nantennas\times\Ncoeffs}$ represents the channel between the BS and RIS, and $\mathbf{G}\triangleq[\mathbf{g}_{1}\,\mathbf{g}_{2}\,\cdots\,\mathbf{g}_{K}]\in\mySet{C}^{\Nantennas\times\Nusers}$ is the direct channel matrix between the BS and the UTs ($\mathbf{g}_{k}\in\mySet{C}^{\Ncoeffs\times1}$ represents the channel for the $k$-th UT). In addition, $\mathbf{n}_t\in\mySet{C}^{\Nantennas}$ is the Additive White Gaussian Noise (AWGN) vector, which is usually modeled as having zero-mean elements and covariance matrix $\sigma^2\mathbf{I}_{\Nantennas}$.

\vspace{-0.2cm}
\subsection{Problem Formulation}\label{sebsec:problem formulation}
The \ac{bs} uses the baseband received signal vector $\myVec{y}_t$, i.e., the channel output in \eqref{eq:channel_output}, along with the prior knowledge of the pilot symbols $\myVec{s}_t$, to train its \ac{dnn}-based detector (i.e., multi-antenna RX) and decide the \ac{ris} phase configuration vector $\boldsymbol{\phi}$. Consequently, by letting $ \myParams\in \mySet{C}^{M\times1}$ denote the weights of the \ac{bs}'s DNN-based symbol detector, our goal is to jointly design $\myParams$ and $\boldsymbol{\phi}$ to minimize the \ac{ber}. By representing the DNN operation as the mapping $\psi_{\myParams}(\cdot):\mySet{C}^{\Nantennas\times1\mapsto} \mySet{S}^{\Nusers\times1}$, we seek to minimize the following objective function:  
	 	 \begin{equation}
	 \label{eqn:Objective}
	 \Objective(\myParams ;\boldsymbol{\phi}) =  \E\left\{\left\|\myVec{s}_t - \psi_{\myParams}\left( f_{\boldsymbol{\phi}}\left( \myVec{s}_t\right)  \right)   \right\|_0 \right\}. 
	 \end{equation} 
Obviously, this objective approaches its minimal value of $0$ when the DNN operation mimics the inverse function of $f_{\boldsymbol{\phi}}\left(\cdot\right)$. The \ac{ber} captures the performance of the considered communication system, which is actually determined by the \ac{dnn}-based RX and the \ac{ris} phase configuration, i.e., $\myParams$ and $\boldsymbol{\phi}$. For the DNN design, we need periodic pilots to optimize  $\myParams$ and $\boldsymbol{\phi}$ to gradually improve the system performance. 
	 
	\vspace{-0.2cm}
	\section{Joint Multi-User RX and RIS Design}
	\label{sec:jointly-learned ris}
	\vspace{-0.1cm}
In this section, we first present our approach for optimizing the considered design objective, and then describe the \ac{dnn}-based RX structure together with the adopted training scheme. We also present our method for jointly optimizing the \ac{dnn}-based multi-antenna RX and the \ac{ris} phase configuration.

\vspace{-0.2cm}
\subsection{Proposed Optimization Approach}\label{proposed approach}
\begin{figure}
		\centering
		{\includefig{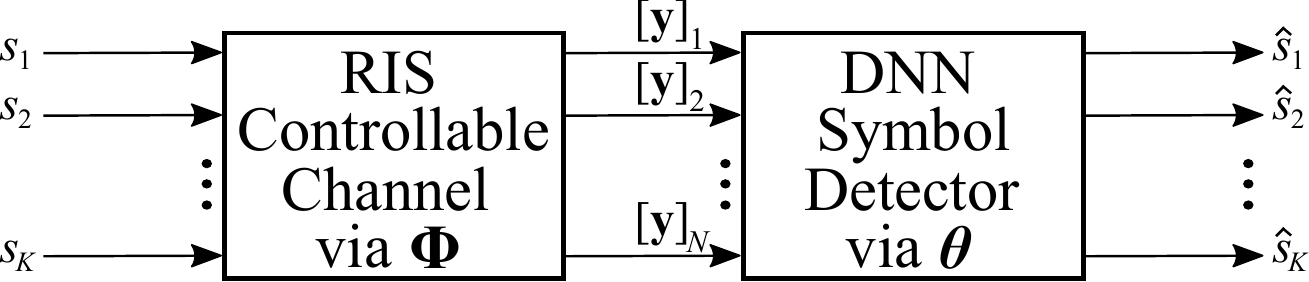}} 
		\caption{The phase configurations of the RIS unit elements considered as hyperparameters in the proposed ML-based approach for the joint design of the BS's multi-antenna RX and the RIS; $\hat{s}_k$ is the estimate for $s_k$.}
		\label{fig:EqModel}	 
	\end{figure}

We seek to jointly adapt the parameters of the proposed DNN-based RX, parameterized by $\myParams$, along with the \ac{ris} phase configuration vector $\boldsymbol{\phi}$, without explicit  knowledge of the channel input-output relationship. The lack of such CSI  renders conventional \ac{ml} optimization algorithms, e.g., based on gradient methods and backpropagation \cite{huang2019spawc, Samarakoon_2020}, infeasible. To tackle this challenge, we note that the \ac{ris} parameters effectively modify the channel conditions, and can thus be considered as {\em hyperparameters}. Those parameters affect the learning process, but are not directly learned in it, i.e., by the DNN-based symbol detector, as illustrated in Fig. \ref{fig:EqModel}. This formulation accounts for the fact that the overall task of the system is to recover the transmitted symbols, which is carried out using the DNN-based detector. Modeling the  elements of $\boldsymbol{\phi}$ as hyperparameters motivates the application of BO for tuning them along with the detector \cite{yu2020hyper}.

Conventional \ac{ml} parameters, e.g., the weights of a neural network which control its mapping, are learned during the training process. Hyperparameters are parameters of \ac{ml} algorithms that control the model's class, e.g., the network architecture \cite{vinyals2016matching}, or the learning process, e.g., the learning rate \cite{maclaurin2015gradient} and the optimization rule \cite{wichrowska2017learned}. They can be either chosen from a discrete set or from a continuous range, and the space which the hyperparameters are chosen from is referred to, henceforth, as the {\em search space}. The main difference between learning the parameters of an \ac{ml} model and optimizing its hyperparameters follows from the fact that: while one can commonly compute the gradient of a model's loss function with respect to its parameters, evaluating the gradients with respect to the hyperparmeters is often infeasible. Consequently, gradient-based methods, which are the leading workhorse in training DNN parameters \cite[Ch. 6]{goodfellow2016deep}, are often not applicable. 
	
It is evident from \eqref{eqn:Objective} that the expected BER in the objective function $\Objective(\myParams ;\boldsymbol{\phi})$ cannot be computed without prior knowledge of the underlying statistical model, and as such, it must be approximated using its empirical form. This requires the transmission of $\Ntraining$ pilot symbols, i.e., a labeled set of the form $\{\myVec{s}_t, \myVec{y}_t\}_{t=1}^{\Ntraining}$ in which $\myVec{y}_t$ is a realization of $f_{\boldsymbol{\phi}}\left( \myVec{s}_t\right)$. Furthermore, to facilitate the optimization of the DNN as a set of classifiers, we consider that the output of the \ac{dnn}-based RX models an estimate of the conditional distribution of each of the transmitted symbols, i.e., $\psi_{\myParams}(\cdot)$ comprises of $K$ vectors of size $M$, each representing an estimate of the conditional probability of a single symbol. In this case, the $\ell_0$-norm in \eqref{eqn:Objective} can be replaced with the cross-enrtopy loss function. By letting  $\psi_{\myParams}(\cdot, \alpha)_k$ represent the estimated probability mass function of the $k$-th symbol evaluated at realization $\alpha \in \mySet{S}$, the resulting empirical cross-entropy loss is given by:
	\begin{equation}
	 \EmpObjective(\myParams ; \boldsymbol{\phi}) =  \sum_{t=1}^{\Ntraining} \sum_{k=1}^{\Nusers} - \log \psi_{\myParams}(\myVec{y}_t, [\myVec{s}_t]_k)_k.
	 \label{eqn:EmpObjective}
	\end{equation}
Using the latter formulation of the empirical error, one can apply the following two-stage iterative procedure for the parameters of the BS's DNN-based RX and those of the RIS phase configuration at each iteration index $i$:
	\begin{align}
	\myParams_{i+1} &= {\arg\min}_{\myParams} 	 \EmpObjective(\myParams ; \boldsymbol{\phi}_i), \label{eqn:ParamOpt}\\
		\boldsymbol{\phi}_{i+1} &= {\arg\min}_{\boldsymbol{\phi}} 	 \EmpObjective(\myParams_{i+1} ; \boldsymbol{\phi}). \label{eqn:CoeffOpt}
	\end{align}
	
We note that, at each iteration $i$ in the above approach, the transmission of additional $\Ntraining$ pilots is required. Recall that, at each time $\boldsymbol{\phi}$ is modified, a channel input-output function $f_{\boldsymbol{\phi}}\left(\cdot\right)$ is generated, and a new training set $\{\myVec{s}_t, \myVec{y}_t\}_{t=1}^{\Ntraining}$ is required. Furthermore, while the optimization of the \ac{dnn} parameters $\myParams$ in \eqref{eqn:ParamOpt} can be carried out using conventional gradient-based methods, updating $\boldsymbol{\phi}$ in \eqref{eqn:CoeffOpt} involves complex optimization with an objective function which is expensive to compute. This follows due to the fact that the hyperparameters $\boldsymbol{\phi}$ affect the channel mapping $f_{\boldsymbol{\phi}}\left(\cdot\right)$, hence, the parameterization of  $f_{\boldsymbol{\phi}}\left(\cdot\right)$ is not explicit, i.e., it is not a known \ac{dnn} architecture as the symbol detector is, and thus the RX has no direct access to it. This indicates that the usage of Bayesian hyperparameter optimization techniques \cite{brochu2010tutorial} for updating $\boldsymbol{\phi}$ should be suitable for the problem at hand, and can be simple to apply using existing BO toolboxes, e.g.,  \cite{balandat2019botorch}. Therefore, the optimization of the objective actually accounts for the optimization of two parameters, i.e., $\myParams$ and $\boldsymbol{\phi}$.

\begin{figure*}
	\subfigure[]{
		\label{fig:deepsic}
		{\includegraphics[width=0.59\textwidth]{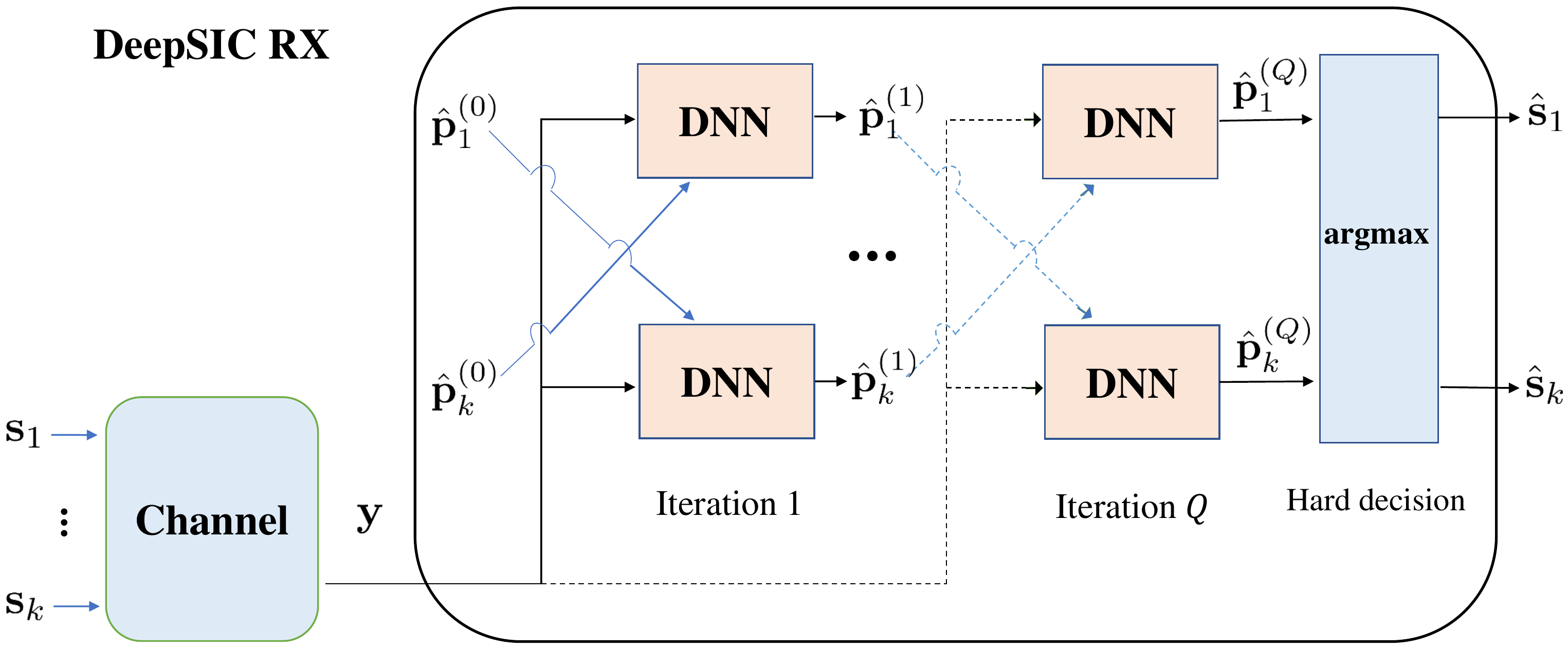}\vspace{-0.2cm}}}
	\subfigure[]{
		\label{fig:dnnarc}
		{\includegraphics[width=0.40\textwidth]{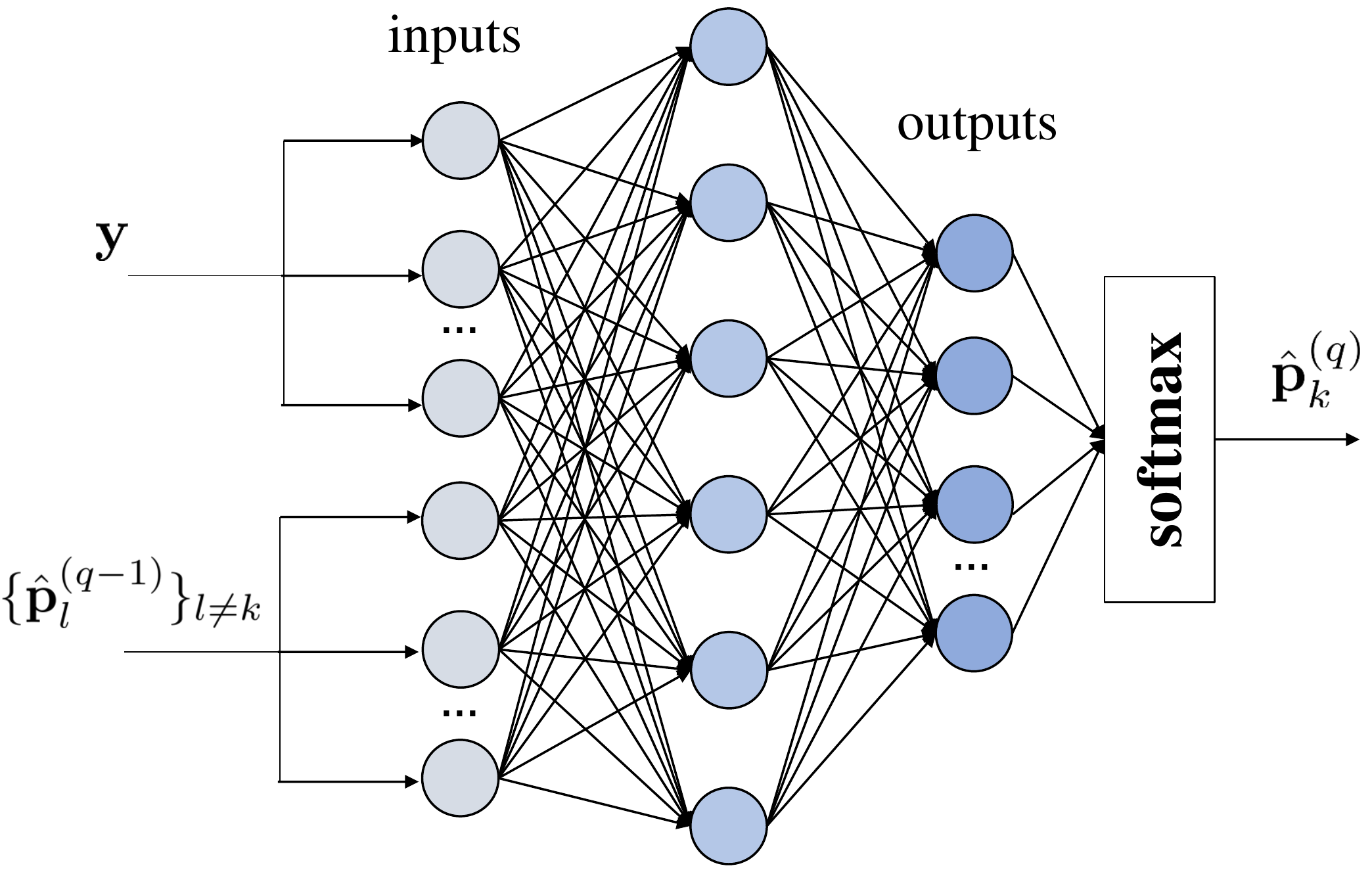}\vspace{-0.2cm}}
	}
	\caption{Structure of the overall DNN-based multi-user RX. (a) DeepSIC-RX illustration; (b) Detailed structure of each DNN in (a).}\vspace{-0.4cm}
\end{figure*}


\vspace{-0.2cm}
\subsection{Multi-User Receiver Model} \label{subsec:receiver model}
We adopt the DeepSIC RX \cite{shlezinger2019deepSIC}, which is a DNN-based soft RX implementing the traditional iterative interference cancellation method, based on channel modeling by means of deep learning, and expands to channel-model-independent implementations. 

\subsubsection{Receiver Architecture}
The RX uses an iterative fashion to achieve interference cancellation. To this end, the symbols transmitted by other UTs are regarded as interference symbols for the $k$-th UT. The detector operates iteratively: in every iteration $ q\in \{1,2,\dots, Q\} \triangleq \mySet{Q} $, an estimate of the conditional distribution, denoted by $ \myVec{p}_k^{(q)}$, of $ \myVec{s}_k $ for a given channel output $ \myVec{y} $ is generated for each UT $ k\in \mySet{K} \triangleq \{1,2, \dots , K\} $  using the corresponding estimate of the interference symbols
$ \{\myVec{s}_l\}_{l\neq k} $ obtained in the previous iteration. Here, we denote $ \myVec{s}_k $ as the symbols transmitted by the $k$-th UT at each time instance. The purpose of interference cancellation is achieved through continuous iteration, and the outputs of the \ac{dnn} at the last iteration are used for decoding in a hard decision manner. The whole process is illustrated in the Fig.~\ref{fig:deepsic}. The \ac{dnn} of each UT in the entire structure can be viewed as a building block and its output is the conditional probability of each UT's symbols. Therefore, it can be seen as a classifier that is agnostic of the channel model.
The output for the $k$-th building block of $q$-th iteration is $ \hat{\myVec{p}}_k^{(q)} $, which is the estimated conditional probability of $\myVec{s}_k $ given $ \myVec{y}$ based on $ \{\hat{\myVec{p}}_k^{(q-1)}\}_{l \neq k}$.

The structure for the \ac{dnn} for each UT's symbol is shown in Fig.~ \ref{fig:dnnarc}. Each soft estimate is produced using a multi-layer fully connected structure with softmax output layer. For simplicity, we illustrate the \ac{dnn} architecture with real-valued channels, as the considered channel model with complex values can be represented by real vectors of an extended dimension.
Since we use classification \acp{dnn} for soft-esitmates, the number of neurons at their output layers depends on the size $M$ of the constellation set. The inputs of each \ac{dnn} includes the inputs at the BS's $N$ receive antennas, and the conditional probabilities  $\{\hat{\myVec{p}}_k^{(q-1)}\}_{l \neq k}$ of the interfering symbols from the previous iteration. Thus, the number of input neurons is $N+(K-1)(M-1)$. 
Finally, the initial conditional probabilities for each UT are set to the uniform distribution, i.e., $ \{{\hat{\myVec{p}}_k^{(0)}}\}_{k=1}^K= M^{-1}$.

\subsubsection{Receiver Training}
In order to make the RX realize reliable symbol detection, each \ac{dnn} needs to be properly trained. We use the data set of $N_{\rm tr}$ pairs of channel inputs and corresponding outputs $ \{{\tilde{\myVec{s}}}_{j},{\tilde{\myVec{y}}}_{j} \}_{j=1}^{N_{\rm tr}}$
to train the RX in a sequential training fashion. Note that the \ac{dnn} building blocks do not depend on the iteration index value, 
and the input of each \ac{dnn} depends on the channel output $ \myVec{y} $ and the output $\{{\myVec{p}}_{l}^{(q-1)}\}_{l\neq k}  $ of the  trained \ac{dnn} in the previous iteration. This indicates that the \acp{dnn} can be trained separately in a sequential manner to minimize the cross-entropy loss. As such, we can train each \ac{dnn} with a small number of samples. By letting $ \myParams_k^{(q)} $ represent the parameters of the $k$-th \ac{dnn} at iteration $q$
 and writing $ \hat{\myVec{p}}_{k}^{(q)}({\myVec{y}},\{\hat{\myVec{p}}_{l}^{(q-1)}\}_{l\neq k} ,\alpha;\myParams_{k}^{(q)}) $ as the entry of $ \hat{\myVec{p}}_k^{(q)}$ corresponding to $\alpha \in \mySet{S} $, when the parameters and inputs of \ac{dnn} are $ \myParams_k^{(q)} $ and $ ({\myVec{y}},\{\hat{\myVec{p}}_{l}^{(q-1)}\}_{l\neq k})$, respectively, we can re-express the empirical cross-entropy loss in \eqref{eqn:ParamOpt} as: 
\begin{equation} \label{eq:cross}
	\mySet{L}(\myParams_{k}^{(q)})=
	\frac{1}{N_{\rm tr}} \sum_{j=1}^{N_{\rm tr}}-\log \hat{\myVec{p}}_{k}^{(q)}(\tilde{\myVec{y}}_{j},\{\hat{\myVec{p}}_{j, l}^{(q-1)}\}_{l\neq k} ,(\tilde{\myVec{s}}_{j})_k; \myParams_{k}^{(q)}),
\end{equation}
where $\{\hat{\myVec{p}}_{j, l}^{(q-1)}\}_{l\neq k}$ represents the estimated conditional probabilities at the previous iteration associated with $ \tilde{\myVec{y}}_{j} $. The sequential training method from \cite{shlezinger2019deepSIC}  is summarized in Algorithm~\ref{algorithm1}.

\begin{algorithm}[!t] \label{algorithm1}
	\SetKwInOut{Input}{Input}
	\SetKwInOut{Output}{Output}
	\caption{Sequential Training for the DeepSIC RX}
	\Input{Training data set $ \{{\tilde{\myVec{s}}}_{j},{\tilde{\myVec{y}}}_{j} \}_{j=1}^{N_{\rm tr}}$, $Q$, and $\mathcal{K}$.}
	\Output{Trained \ac{dnn} parameters $\myParams=\{\myParams_k^{(q)}\}_{k,q=1}^{K,Q}$.}

	\textbf{Initialization:} Set $ q=1 $, $ \{\hat{\myVec{p}}_k^{(0)}\}_{k=1}^K = M^{-1} $
	and  $ \hat{\myVec{p}}_{j,k}^{(0)}=\hat{\myVec{p}}_k^{(0)} $
	for every $ k \in \mySet{K}$, and $ j \in \{1,2,\dots,N_{\rm tr}\} $.   \\
	
	\While{$ q \leq Q  $}
	{\For{ $ k \in \mySet{K} $}
		{ Set the \ac{dnn} parameters to $ \myParams_k^{(q)} $. \\
		  Train the \ac{dnn} to minimize (\ref{eq:cross}).  \\
		  Feed$ \{\tilde{\myVec{y}}_j,\{\hat{\myVec{p}}_{j,l}^{(q-1)}\}_{l\neq k}\}_{j=1}^{N_{\rm tr}} $ to the trained \ac{dnn} to obtain $\{\hat{\myVec{p}}_{j,k}^{(q)}\}_{j=1}^{N_{\rm tr}} $ .\\
		  Set  $  q = q+1 $ . \\
		}
	}
\end{algorithm}

\vspace{-0.2cm}
\subsection{Proposed Joint RX and RIS Learning Method}


In this section, we first discuss the application of the \ac{bo} framework to our RIS-empowered multi-user MIMO system design objective, and then present a method to jointly optimize the \ac{dnn}-based multi-antenna RX and the RIS phase configuration.
 
\subsubsection{RIS Configuration via Bayesian Optimization} \label{ris optimization}  
Let $g(\cdot)$ represent the mapping between the reflection coefficients of the \ac{ris} and the RX output, and ee focus on finding $\boldsymbol{\phi}^*\triangleq{\arg\max}_{\boldsymbol{\phi}}g(\boldsymbol{\phi}) $ (or equivalently, one can solve for ${\arg\min}_{\boldsymbol{\phi}}{-g(\boldsymbol{\phi})}$). Since it's hard to acquire knowledge for $g(\cdot)$ (i.e., it's a black box function), traditional gradient methods cannot be applied for learning it. For such cases, the BO formulation can be used which consists, in general, of the following two components: \textit{i}) a \textit{surrogate model} (the most commonly used is the \ac{gp}) to incorporate prior beliefs about the objective function; and \textit{ii}) an \textit{acquisition function} that directs sampling to areas where an improvement over the current best observation is likely.

Let $ \boldsymbol{\phi}_i $ denote the $i$-th sample RIS phase configuration and $ g(\boldsymbol{\phi}_i) $ the output observation at the RX of the unknown objective function at the point $\boldsymbol{\phi}_i$. We also use the set notation $\mySet{D}_{1:n}\triangleq\{\boldsymbol{\phi}_{1:n},\myVec{g}_{1:n}\}$ to represent the observed data pairs; the subscript is used to denote sequences of data, i.e., $ \boldsymbol{\phi}_{1:n}\triangleq[\boldsymbol{\phi}_1,\boldsymbol{\phi}_2,\dots,\boldsymbol{\phi}_n]$ and $ \myVec{g}_{1:n}\triangleq[g(\boldsymbol{\phi}_1),g(\boldsymbol{\phi}_2),\dots,g(\boldsymbol{\phi}_n)]$. Following the Bayesian framework, the observation $\myVec{g}_{1:n}$ can be considered to be drawn randomly from some prior probability distribution. Usually, GP considers the multivariate normal as the prior distribution, having as unknown parameters its $n$-element mean value vector $\boldsymbol{\mu}$ and its $n\times n$ covariance matrix $\mathbf{K}$; the latter is also referred as \textit{kernel} denoted  $k$ . A popular choice of the kernel is squared exponential function,
\begin{equation}
k\left(\boldsymbol{\phi}_i, \boldsymbol{\phi}_j\right)=\exp \left(-\frac{1}{2}\left\|\boldsymbol{\phi}_i-\boldsymbol{\phi}_j\right\|^{2}\right)
\end{equation}
Hence, the prior distribution on $ \myVec{g}_{1:n} $ is mathematically expressed as:
\begin{equation} \label{normal}
	 \myVec{g}_{1:n} \sim \mySet{N}\left(\boldsymbol{\mu}\left(\boldsymbol{\phi}_{1:n}\right), \mathbf{K}\left(\boldsymbol{\phi}_{1:n}\right)\right),
\end{equation}
where $\boldsymbol{\mu}(\boldsymbol{\phi}_{1:n})$ and $\mathbf{K}\left(\boldsymbol{\phi}_{1:n}\right)$ include the sample mean values and the covariance matrix for the observation $\boldsymbol{\phi}_{1:n}$; for simplicity, we will henceforth omit the dependence of these parameters on $\boldsymbol{\phi}_{1:n}$, although implied, and write $\boldsymbol{\mu}_{1:n}$ and $\mathbf{K}_{1:n}$. 
The latter implies that the observation set $\mySet{D}_{1:n}$ is used to compute $\myVec{g}_{1:n}$, thus providing an estimate for the unknown mapping $g(\cdot)$. The BO framework can be then used to predict the next value $g_{n+1}$ at an arbitrary point $\boldsymbol{\phi}_{n+1} $, hence refining $g(\cdot)$ estimation. According to the properties of GPs, $\myVec{g}_{1:n}$ and $g_{n+1}$ (which denotes a new observation at the RX output as a response to a new RIS configuration $\boldsymbol{\phi}_{n+1}$) follow the bivariate Gaussian distribution. By using the notation $\tilde{\mathbf{K}}_{1:n,n+1}$
to denote the covariance matrix of that distribution and applying the Bayes' rule, the prediction for the conditional distribution of $g_{n+1}$, given the observations $\mySet{D}_{1:n}$ and the RIS phase configuration $\boldsymbol{\phi}_{n+1}$, is expressed as:
\begin{equation}
	\begin{aligned}
	g_{n+1} \mid \mySet{D}_{1:n} ,\boldsymbol{\phi}_{n+1} & \sim \mySet{N}\left(\mu_n(\boldsymbol{\phi}_{n+1}), \sigma_n^{2}(\boldsymbol{\phi}_{n+1})\right),
	\end{aligned}
\end{equation}
where the mean and variance are respectively given by:
\begin{equation}
	\begin{aligned}
	\mu_n(\boldsymbol{\phi}_{n+1}) &\triangleq \tilde{\mathbf{K}}_{1:n,n+1}^{\mathrm{T}}\myMat{K}^{-1}(\myVec{g}_{1:n}-\boldsymbol{\mu}_{1:n}
	) +\mu(\boldsymbol{\phi}_{n+1}),             \\
	\sigma_n^{2}(\boldsymbol{\phi}_{n+1}) &\triangleq
	k(\boldsymbol{\phi}_{n+1},\boldsymbol{\phi}_{n+1}) \!-\!\tilde{\mathbf{K}}_{1:n,n\!+\!1}^{\mathrm{T}}\myMat{K}_{1:n}^{-1}\tilde{\mathbf{K}}_{1:n,n\!+\!1}.
	\end{aligned}
\end{equation}

The latter conditional distribution is called the \textit{posterior} probability distribution. Its mean implies the actual prediction and its variance represents the value of uncertainty. The previous process constitutes the first part of the BO framework. It is a sequential iterative process of obtaining the posterior from the prior and constantly updating the prior. In other words, after modeling the unknown mapping $g(\cdot)$ with \ac{gp}, we can learn its posterior distribution at any sampling point $\boldsymbol{\phi}_i$, which helps us to find its optimal value. When a new observation is collected, we add it to the existing data set to form an enlarged new set, and then, update the prior. Thus, by expanding the observation set, i.e., as $n$ increases for $\mySet{D}_{1:n}$, the fitted function will approximate the unknown mapping. 

The problem now is how to choose the next sampling point efficiently, so that we can find the optimal value with the least number of iterations. This problem can be solved by the second component of BO, i.e., the \textit{acquisition function} which helps to guide the search. In general, every next sample in BO is obtained by maximizing the acquisition function. That is, we want to sample $g$ at $ {\arg\max}_{\boldsymbol{\phi}}p(\boldsymbol{\phi}|\mySet{D}) $, where $p(\cdot)$ represents the generic function of the acquisition function. In this paper, we deploy the popular analytic acquisition function of the \ac{ei} to find the next sample point for the Bayes' rule. The definition of improvement denoted I for an input $\boldsymbol{\phi}$ is:
\begin{equation} \label{ei}
	\operatorname{I}({\boldsymbol{\phi}})\triangleq\max \left(g_{n+1}({\boldsymbol{\phi}})-g\left({\boldsymbol{\phi}}^{+}\right), 0\right),
\end{equation}
where $g({\boldsymbol{\phi}}^{+})$ denotes the currently observed best value and $g_{n+1}({\boldsymbol{\phi}})$ is the mapping output at the next sample point. This function is intuitively easy to understand: it is positive when the prediction is greater than the best observation so far; on the contrary, it is set to zero. Therefore, the second step in (\ref{eqn:CoeffOpt}) of our iterative procedure can be determined by optimizing the \ac{ei} acquisition function, i.e.:
\begin{equation}\label{query point}
	\boldsymbol{\phi}_{i+1}\!=\!{\arg\max}_{\boldsymbol{\phi}}\, 
	\mathbb{E}\!\left\{\max \left(g_{n+1}(\boldsymbol{\phi})\!-\!g({\boldsymbol{\phi}}^{+}), 0\right)\!\!\mid\!\!\mySet{D}_{1:n}\right\}\!.
\end{equation}
It is noted that this function can be evaluated analytically under the GP model\cite{Jones1998}. This means that the gradients are available and we can easily optimize this function using relevant toolboxes \cite{goodfellow2016deep}.

\subsubsection{Joint RX and RIS Optimization}\label{joint optimization}
As shown in \eqref{eqn:ParamOpt} and \eqref{eqn:CoeffOpt}, the proposed joint design approach includes the following sequential iterative procedure: the \ac{dnn}-based multi-antenna RX is first optimized for multi-user decoding for a given RIS configuration, and then, the \ac{ris} optimization is performed via \ac{bo} to further reduce the \ac{ber} performance. We consider sending $T$ pilot symbols at each iteration index $t$ and for a given RIS phase configuration to train the DeepSIC RX, with the intention to optimize the parameters $\myParams$. When the \ac{ris} is optimized and $\boldsymbol{\phi}_{t+1}$ for the next iteration is obtained, the wireless propagation channel changes. This means that the new \ac{ris} configuration $\boldsymbol{\phi}_{t+1}$ will generate a new function $f_{\boldsymbol{\phi}_{t+1}}(\cdot)$ for the input-output relationship of the wireless channel. Consequently, we need to resend pilots to estimate this channel and retrain the DeepSIC RX for multi-user symbol detection. It is noted that the entire optimization process needs to be carried out alternately in a sequential manner. The steps of the proposed approach are included in Algorithm \ref{algorithm2}.

In the proposed algorithm, we use the set $\{\myVec{s}_t,\myVec{y}_t\}$ to present the channel input and corresponding output at the iteration index $t$. The set $ \mySet{D}_{1:t}$ includes the data $\{\boldsymbol{\phi}_{1:t},{\rm BER}_{1:t}\}$, with ${\rm BER}_{1:t}$ representing the BER performances of the trained DeepSIC RX up to the time index $t$, which is used for the proposed BO method. The RX training in Step $5$ is implemented with Algorithm \ref{algorithm1}. We also use additional UT data to validate the trained RX, by computing the \ac{ber}. At the end of the algorithmic iterations $N_{\rm bo}$, we select the \ac{ris} phase configuration corresponding to the minimum \ac{ber}, i.e., $\boldsymbol{\phi}^*$, as shown in Step $10$. It is worth noting that after obtaining the optimal $\boldsymbol{\phi}^*$, we still choose to transmit pilots, in particular the $\myVec{s}_1$ sent during the first iteration, to form the corresponding channel output $\myVec{y}^*$ for retraining the DeepSIC RX. In this way, the best parameters of the trained \ac{dnn} and the \ac{ris}, i.e., $\myParams^*$ and  $\boldsymbol{\phi}^*$, are obtained. It is noted that the first algorithmic iteration is considered as the initial reference and the remaining $N_{\rm bo}-1$ iterations are used for the optimization process. At the end of the process, we re-send $\myVec{s}_1$ to verify whether the \ac{ris} parameters have been optimized. In addition, at each iteration index $t$, the $T$ transmitted pilots are chosen to be different in order to improve the generalization capability of the system. Also, the initial \ac{ris} configuration $\boldsymbol{\phi}_1$ is randomly chosen since there is no CSI available at the system. 

\begin{algorithm}[!t]\label{algorithm2}
	
	\caption{ML-Based Joint RX and RIS Design}
	\textbf{Initialization:} Generate the \ac{ris} phase configuration $\boldsymbol{\phi}_1$ and the \ac{dnn} parameters $\myParams_1$, and initialize the observation set as $\mySet{D}_{1:1}=\{\emptyset\}$. \\
	\For{$t=1,2,\ldots,N_{\rm bo}$}
	{
		The UTs make use of the $T$ pilots ${\myVec{s}_t}$ and the BS sets as $\myParams_t$ the parameters of its \ac{dnn}-based RX. \\
		The RIS realizes $\boldsymbol{\phi}_t$ and the data set $(\myVec{s}_t,\myVec{y}_t)$ is generated at the RX input after UTs' transmission.    \\		
		The RX is trained to obtain $\myParams_{t+1}$ and ${\rm BER}_t$.   \\ 
		The new data set $(\boldsymbol{\phi}_t,{\rm BER}_t)$ is generated and appended to the existing set $\mySet{D}_{1:t-1}$.  \\
		The data set $\mySet{D}_{1:t}=\{\boldsymbol{\phi}_{1:t},{\rm BER}_{1:t}\}$ is formulated to estimate $g(\cdot)$ via the BO approach. \\
		Solve \eqref{query point} to get the next sample $\boldsymbol{\phi}_{t+1}$.
	}
	Set $\boldsymbol{\phi}^*=\boldsymbol{\phi}_{t^*}$ where $t^*={\arg\min_{t}}{\rm BER}_{1:N_{\rm bo}}$. \\
	Run Steps $3$--$5$ with $\myVec{s}_1$ and $\boldsymbol{\phi}^*$ to obtain $\myParams^*$ for the  RX.\\ 
	\textbf{Output:} The joint design $\boldsymbol{\phi}^*$ and $\myParams^*$.
\end{algorithm}



\vspace{-0.2cm}
\subsection{Discussion}
The proposed joint RX and RIS design approach is quite flexible. During the \ac{bo} process, we model the mapping between the \ac{ris} and RX, which is independent of the specific RX structure, i.e., the RX is relatively independent. This greatly improves the scalability of the framework. The deployed DeepSIC RX \cite{shlezinger2019deepSIC} is suitable for multi-user \ac{mimo} scenarios, implementing a model-based ML algorithm, but also extending it to a channel-model-independent realizations; this is suitable for both linear and non-linear channels. Each \ac{dnn} building block in DeepSIC is trained separately, thus, its complexity increases linearly with the DeepSIC of UTs; this is very beneficial for \ac{mimo} systems. In addition, DeepSIC allows to complete the training process with a small number of training sets, i.e., we can send few pilots for training at each iteration. The computational complexity of the \ac{bo} approach to optimize the \ac{ris} is acceptable.
Since we consider probabilistic modeling, the problem of optimizing a black box function is transformed into that of optimizing an acquisition function, which has reduced computational complexity.

The training of the DeepSIC RX is carried out in an iterative way, which means that the improvement in the BER performance is at the cost of increasing training time. Note that in Algorithm~\ref{algorithm2}, the DeepSIC RX needs to be retrained at each iteration. This happens because we need to continuously obtain the data used to fit the target mapping in the \ac{bo} formulation, and the system performance can be gradually improved as the amount of data increases. Therefore, the BER improvement in the propsoed two-stage iteration is at the cost of increasing time complexity. Actually, the \ac{bo} method is typically applied to adapt a limited amount of parameters. One can extend it to high dimensions \cite{han2020high}; we leave this for a future extension. 


\vspace{-0.4cm}
\section{Numerical Results}
\label{sec:simulation study}
\vspace{-0.1cm} 
\begin{figure*}
	\subfigure[]{
		\label{fig:simulation}
		{\includegraphics[width=0.92\columnwidth]{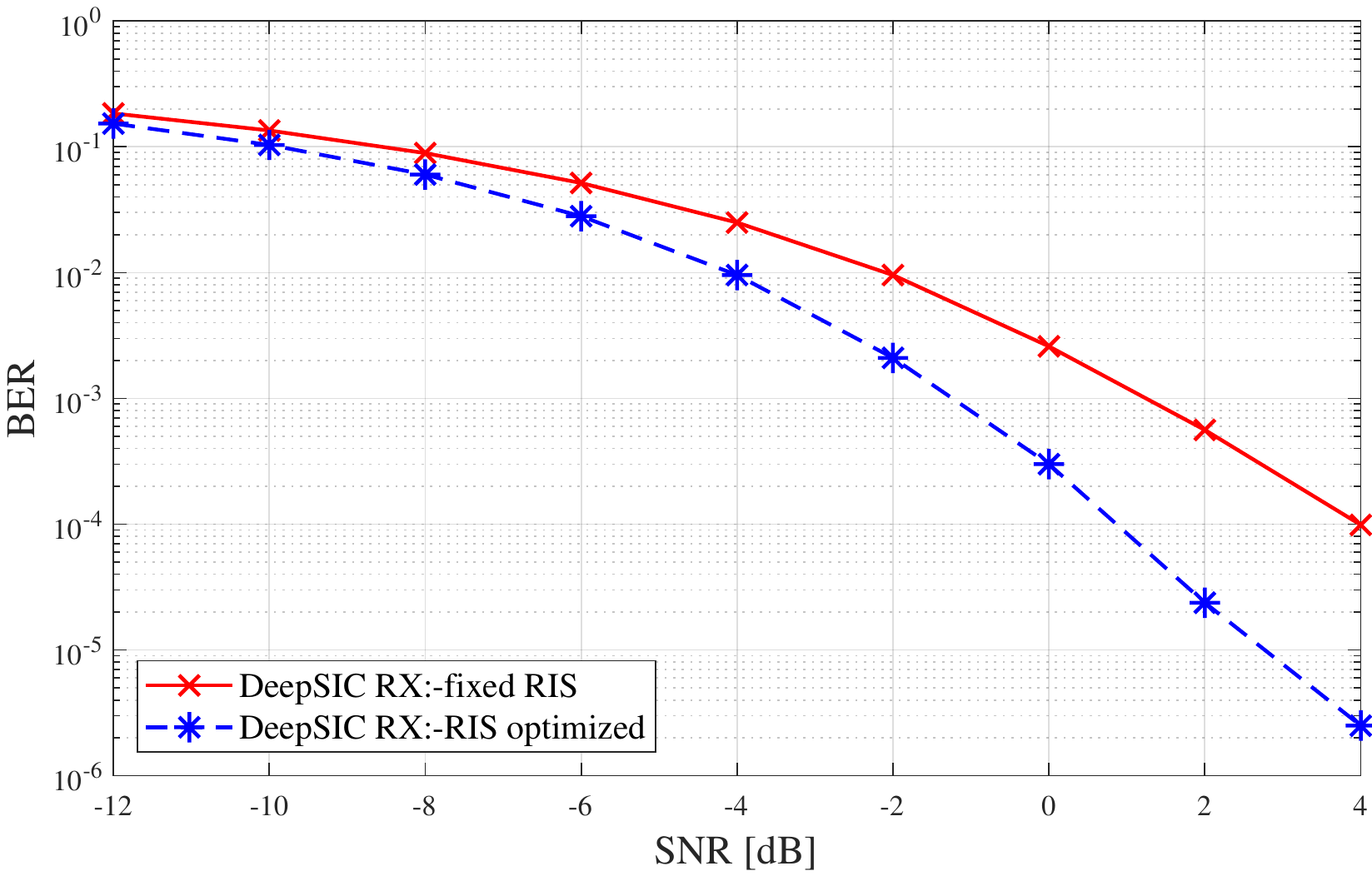}\vspace{-0.2cm}}}
	\subfigure[]{
		\label{fig:random-BO}
		{\includegraphics[width=0.95\columnwidth]{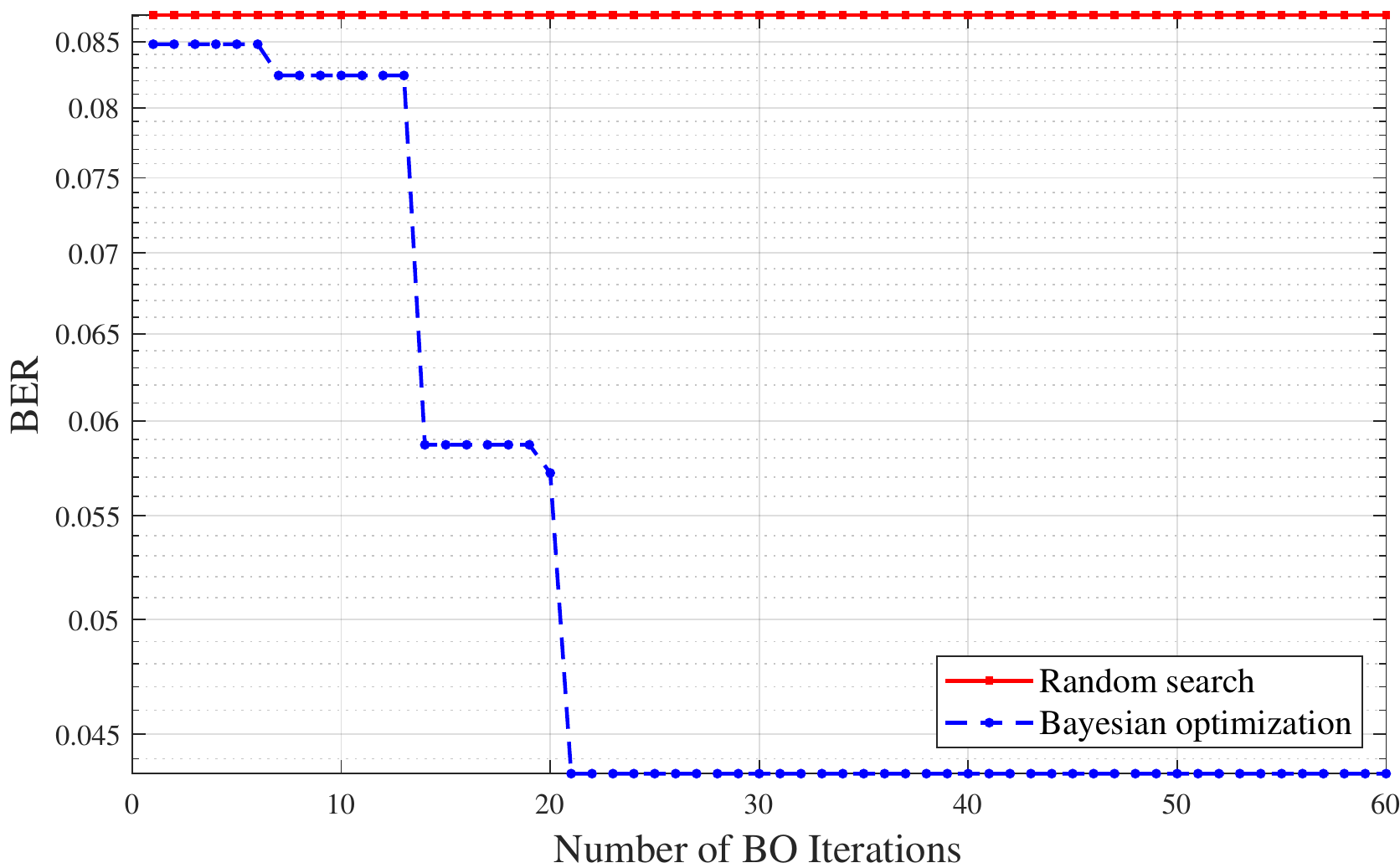}\vspace{-0.2cm}}
	}
	\centering
	\caption{The BER performance of the proposed ML-based joint RX and RIS optimization: (a) versus the SNR in dB; (b): versus the iteration number $N_{\rm bo}$.}\vspace{-0.4cm}
\end{figure*}

\subsection{Simulation Parameters}\label{subsec:simualtion settings}
The scenario we consider is the uplink of a \ac{mimo} communication system assisted by an \ac{ris}. There exist $K=5$ UTs in the system, the BS has $N=5$ antenna elements, and the \ac{ris} consists of $P=18$. The considered channel model follows \eqref{eq:channel model}, where the channel matrices $\mathbf{H}_1$ and $\mathbf{G}$ follow the Rayleigh distribution, while $\mathbf{H}_2$ is a Rician channel. As such, we have modeled each column of $\mathbf{H}_1$ and $\mathbf{G}$ as $\mathbf{h}_{1,k}=\gamma\tilde{\mathbf{h}}_{1,k}$ and $\mathbf{g}_{k}=\gamma\tilde{\mathbf{g}}_{k}$, respectively, where $\gamma$ is the passloss factor and $ \tilde{\mathbf{h}}_{1,k},\gamma\tilde{\mathbf{g}}_{k} \sim \mySet{CN}(0,\mathbf{I}_P)$. The channel matrix $\mathbf{H}_2$ was modeled as follows:
\begin{equation}
	\mathbf{H}_2=\beta\left(\sqrt{\frac{\kappa}{1+\kappa}} \tilde{\mathbf{H}}_2^{\mathrm{LOS}}+\sqrt{\frac{1}{1+\kappa}} \tilde{\mathbf{H}}_2^{\mathrm{NLOS}}\right),
\end{equation}
where $ \beta $ and $ \kappa $ denote the pathloss and Rician factors, respectively. The superscripts 'LOS' and 'NLOS' represent the LOS and non-LOS components of the channel. These channel components in the previous expression were both modeled as standard Gaussian channel matrices. In the simulations, we have set the passloss factors $\beta$ and $\gamma$ to be normalized to $1$, and the Rician factor $\kappa $ was set $10$. 

The \ac{dnn} structure of the DeepSIC RX was chosen to have three fully-connected layers: the $ N+(K-1)(M-1)\times60 $ first layer followed by sigmoid activation, the $ 60\times30 $ second layer followed by ReLU activation, and the $ 30\times M $ third layer. The transmitted symbols were randomly generated from the QPSK constellation, i.e., $ \mySet{S}_{qpsk}=\{1+j, -1+j, -1-j, 1-j\} $. Note that this modulation scheme can be represented equivalently by the BPSK constellation, i.e., $ \mySet{S}_{bpsk}=\{-1,1\}$ in the following sence: a $ 5\times5 $ complex channel with QPSK signaling is actually equivalent to a $ 10\times10 $ real channel with BPSK signaling. This transformation can be actually used to simplify simulations. Since the input data $\mathbf{y}$ at the \ac{dnn}-based RX cannot be complex, we have given as inputs the real and imaginary parts of this vector. The number of iterations $Q$ for each \ac{dnn} run was set to $5$ in Algorithm~\ref{algorithm1} and the optimizer used was the ADAM with learning rate $0.01$. We have used Python as the simulation environment and we realized BO via Botorch based on Pytorch, which is a BO toolbox.

\vspace{-0.2cm}
\subsection{BER Performance Results}\label{subsec:simulation rusult}
We first evaluated the \ac{ber} performance versus the \ac{snr}, defined as $ 1/\sigma^2$, for a fixed \ac{ris} configuration in order to solely optimize the DeepSIC RX and verify its performance. Then, we evaluated the BER with the proposed joint DeepSIC RX and RIS optimization, which is described by the two-stage iteration method summarized in Algorithm~2. The results in Fig.~\ref{fig:simulation} were obtained with a training set of $1000$ samples and a testing data with more than $80000$ symbols. The total number of iterations for the BO part were set as $N_{\rm bo}=25$. It can be seen that when \ac{ris} is fixed, the DeepSIC RX can achieve a good BER performance with a small training data set, and as expected, \ac{ber} decreases gradually with the \ac{snr}. Interestingly, when we also optimize the \ac{ris} phase configurations, the \ac{ber} gets further reduced. For example, for both cases to achieve the BER value $10^{-3}$, there will be more than $2$ dB gain after optimizing the \ac{ris}.

In Fig.~\ref{fig:random-BO}, we evaluate the \ac{ber} versus the BO iteration number $N_{\rm bo}$ for the fixed \ac{snr} value $ -8$ dB. It can be concluded that the \ac{ber} decreases with increasing $N_{\rm bo}$. Note that in Algorithm~2, we get the next sample by optimizing the acquisition function. In this subfigure, we also simulated the case where the next sample point (i.e., next RIS configuration) was generated in a random way. Namely, other conditions remained unchanged and we transmitted the same pilots at each iteration, while randomly generating the \ac{ris} phase configurations. This random generation yields a straight BER curve, which means that none of the chosen \ac{ris} configurations improved the BER in the considered $60$ iterations. Recall that, at each iteration in our proposed approach, we need to send pilot signals and train the DeepSIC RX receiver whenever the \ac{ris} configuration changes. This results in training overhead, implying that the \ac{ris} configuration search needs to perform efficiently, improving the BER sequentially. As shown, our BO method is more stable than the random selection strategy, yielding better RIS configurations (in terms of BER) with fewer algorithmic iterations.

\vspace{-0.2cm}
\section{Conclusions}
\label{sec:conclusions}
\vspace{-0.1cm} 
In this paper, we proposed a joint optimization approach for reception and RIS phase configuration targeting reliable symbol detection in the uplink of RIS-aided multi-user MIMO communication systems. We considered a DNN-based RX which was combined with BO in a two-stage alternating optimization algorithm, requiring periodical transmissions of small numbers of pilot symbols for convergence. Our BER performance evaluation results demonstrated the efficacy of the proposed model-agnostic DNN-based learning approach. 

%
\vspace{-0.15cm}
	\bibliographystyle{IEEEtran}
	\bibliography{IEEEabrv,refs}
 
%
\end{document}